\pdfoutput=1

\documentclass[sigconf]{acmart}

 \usepackage{multirow}
\AtBeginDocument{%
  \providecommand\BibTeX{{%
    \normalfont B\kern-0.5em{\scshape i\kern-0.25em b}\kern-0.8em\TeX}}}

\copyrightyear{2020} \acmYear{2020} \setcopyright{acmcopyright}\acmConference[CIKM '20]{Proceedings of the 29th ACMInternational Conference on Information and Knowledge Management}{October19--23, 2020}{Virtual Event, Ireland}\acmBooktitle{Proceedings of the 29th ACM International Conference on Informationand Knowledge Management (CIKM '20), October 19--23, 2020, Virtual Event, Ireland}\acmPrice{15.00}\acmDOI{10.1145/3340531.3412128}\acmISBN{978-1-4503-6859-9/20/10}

\usepackage{hyperref}
\usepackage{cleveref}

\DeclareMathOperator{\x}{\textit{\textbf{x}}}
\DeclareMathOperator{\X}{\textit{\textbf{X}}}
\DeclareMathOperator{\E}{\mathbb{E}}
\newcommand{\p}{\textup{\texttt{+}}} 
\newcommand{\m}{\textup{\texttt{-}}} 

\begin{document}
	\fancyhead{}
	\title{Analysis of Multivariate Scoring Functions for Automatic Unbiased Learning to Rank}
	
	\author{Tao Yang}
	\email{taoyang@cs.utah.edu}
	
	\affiliation{%
		\institution{University of Utah}
		\streetaddress{50 Central Campus Dr.}
		\city{ Salt Lake City}
		\state{Utah}
		\postcode{84112}
	}
	\author{Shikai Fang}
	\email{shikai.fang@utah.edu}
	\affiliation{%
		\institution{University of Utah}
		\streetaddress{50 Central Campus Dr.}
		\city{ Salt Lake City}
		\state{Utah}
		\postcode{84112}
	}
	
	\author{Shibo Li}
	\email{shibo@cs.utah.edu}
	\affiliation{%
		\institution{University of Utah}
		\streetaddress{50 Central Campus Dr.}
		\city{ Salt Lake City}
		\state{Utah}
		\postcode{84112}
	}

	\author{Yulan Wang}
	\email{yulan.wang@utah.edu}
	\affiliation{%
		\institution{University of Utah}
		\streetaddress{50 Central Campus Dr.}
		\city{ Salt Lake City}
		\state{Utah}
		\postcode{84112}
	}
	
	\author{Qingyao Ai}
	\email{aiqy@cs.utah.edu}
	\affiliation{%
		\institution{University of Utah}
		\streetaddress{50 Central Campus Dr.}
		\city{ Salt Lake City}
		\state{Utah}
		\postcode{84112}
	}

	%

	
	\begin{abstract}
		Leveraging biased click data for optimizing learning to rank systems has been a popular approach in information retrieval.
		Because click data is often noisy and biased, 
		a variety of methods have been proposed to construct unbiased learning to rank (ULTR) algorithms for the learning of unbiased ranking models.
		Among them, 
		automatic unbiased learning to rank (AutoULTR) algorithms that jointly learn user bias models (i.e., propensity models) with unbiased rankers have received a lot of attention due to their superior performance and low deployment cost in practice.
		Despite their differences in theories and algorithm design, existing studies on ULTR usually use uni-variate ranking functions to score each document or result independently.
		On the other hand, recent advances in context-aware learning-to-rank models have shown that multivariate scoring functions, which read multiple documents together and predict their ranking scores jointly, are more powerful than uni-variate ranking functions in ranking tasks with human-annotated relevance labels.
		Whether such superior performance would hold in ULTR with noisy data, however, is mostly unknown.
		In this paper,
		we investigate existing multivariate scoring functions and AutoULTR algorithms in theory and prove that permutation invariance is a crucial factor that determines whether a context-aware learning-to-rank model could be applied to existing AutoULTR framework.
		Our experiments with synthetic clicks on two large-scale benchmark datasets show that AutoULTR models with permutation-invariant multivariate scoring functions significantly outperform those with uni-variate scoring functions and permutation-variant multivariate scoring functions.
	\end{abstract}

	\ccsdesc[500]{Information systems~Learning to rank}
	\keywords{Unbiased learning to rank; Multivariate Scoring Function}
	
	
	
	\maketitle
	\vspace{-0.4cm}
	\section{Introduction}
	\vspace{-0.1cm}
	The study of learning to rank with implicit user feedback such as click data has been extensively studied in both academia and industry. Usually, learning to rank directly with implicit user feedback would suffer from the intrinsic noise and propensity in user interaction(e.g., position bias). In recent years, many algorithms have been proposed to find the best model that ranks query-document pairs according to their intrinsic relevance. Among them, unbiased learning to rank (ULTR) algorithms that automatically estimate click bias and construct unbiased ranking models, namely the AutoULTR algorithms, have drawn a lot of attention~\cite{ai2018unbiased,wang2018position}.
	Because they do not need to conduct separate user studies or online experiments to estimate user bias (i.e., the propensity models), most AutoULTR algorithms can easily be deployed on existing IR systems without hurting user experiences.
	
	
	Despite their differences in background theories and algorithm design, previous studies on ULTR usually use uni-variate learning-to-rank models, which score each document independently, for experiments and theoretical analysis.
	Recently, multivariate scoring functions, which take multiple documents as input and jointly predict their ranking scores, have been proved to be more effective than uni-variate scoring function in many learning-to-rank problems~\cite{ai2018learning, pang2020setrank}. 
	By modeling and comparing multiple documents together, multivariate scoring functions naturally capture the local context information and produce state-of-the-art performances on many learning-to-rank benchmarks with human-annotated relevance labels.
	Whether the superior performance of multivariate scoring functions would hold in unbiased learning to rank is still unknown.
	
	To leverage the full power of click data and multivariate scoring functions, we explore the potential of multivariate scoring functions in AutoULTR. Specifically, we investigate the compatibility of DLA~\cite{ai2018unbiased}, an AutoULTR algorithm, with two existing multivariate scoring functions, SetRank~\cite{pang2020setrank}, and DLCM~\cite{ai2018learning}. 
	Our theoretical analysis shows that only a subset of multivariate scoring functions could perform well in ULTR.
	Specifically, we prove that being permutation invariant is a crucial factor determining whether a multivariate scoring function could be applied to the existing AutoULTR framework.
	Experiment results with synthetic clicks on two large-scale publicly available benchmarks showed that the DLA with permutation-invariant multivariate scoring functions significantly outperforms DLA with uni-variate scoring functions and permutation-variant multivariate scoring functions. 
	\vspace{-0.2cm}
	\section{Related work}
	\vspace{-0.1cm}
	AutoULTR, which has the advantage of estimating propensity and relevance simultaneously, has drawn much attention recently. For example, Ai et al. \cite{ai2018unbiased} proposed Dual Learning Algorithm (DLA) based on counterfactual learning, and Wang et al. \cite{wang2018position} proposed a regression-based EM algorithm. On the other hand, for multivariate scoring functions, there are several ways to take multiple documents as input. For example, DLCM \cite{ai2018learning} employs a RNN model to give score, SetRank\cite{pang2020setrank} utilizes self-attention mechanism.
	To the best of our knowledge, however, research on multivariate scoring functions in AutoULTR has not been fully explored, which is exactly the focus of this paper.
	\vspace{-0.2cm}
	\section{PROBLEM FORMULATION}
	\vspace{-0.1cm}
	In this section, we investigate existing AutoULTR algorithms in theory and prove that permutation invariance is a sufficient and necessary condition that determines whether multivariate scoring functions could be applied to existing AutoULTR algorithms. A summary of the notations used in this paper is shown in Table \ref{tab:notation}.
	\begin{table}[t]
		\setlength{\belowcaptionskip}{-10pt}
		\caption{A summary of notations.}
		\vspace{-0.4cm}
		\small
		\def\arraystretch{1.15}
		\begin{tabular}
			{| p{0.05\textwidth} | p{0.38\textwidth}|} \hline
			$ Q, q$ &All possible query $Q$ and a query instance $q \sim P(q)$.   \\\hline
			$S$,$\mathcal{F}$,$\theta$, $E$,$\mathcal{G}$, $\phi$ & A multivariant relevance estimation function $\mathcal{F}$ parameterized by $\theta$  for ranking system $S$ and a propensity estimation function $\mathcal{G}$ parameterized by $\phi$ for propensity model $E$ . \\\hline
			$l$, $\widetilde{l}$,$\widetilde{\mathcal{L}}$ & $l$ is local loss, while $\widetilde{l}$ is unbiased estimation of $l$ and  $\widetilde{\mathcal{L}}$ is the unbiased estimation of global loss.\\\hline
			
			$\pi_{q}$, $d_i$, $i$,  $\x_i$, $\X$, $y$ & A ranked list $\pi_q$ produced by $S$ for $q$, a document $d_i$ with features $\x_i$ on the $i$-th position in $\pi_q$ and its relevance $y$. $\X$ is the feature matrix for whole $\pi_q$. \\\hline
			$\mathbf{o}_{q}$, $\mathbf{r}_{q}$, $\mathbf{c}_{q}$ &Bernoulli variables that represent whether a document is observed ($\mathbf{o}_{q}$), perceived as relevant ($\mathbf{r}_{q}$) and clicked ($\mathbf{c}_{q}$ ).\\\hline
		\end{tabular}\label{tab:notation}
		\vspace{-20pt}
	\end{table}
	\vspace{-0.3cm}
	\subsection{Univariate and Multivariate Scoring}
	In a standard learning-to-rank system, given a specific query $q$ and its associated retrieved document set $D=[d_1,d_2,\dots,d_N]$, a vector $\x_i \in \mathbb{R}^{H}$ can be extracted and used as the feature representation for $d_i$. Let $\pi_q$ be the ranked list for  query $q$. Then there will be a feature matrix for a ranked list $\pi_q$:
	\begin{equation*}
		\X=[\x_1,\x_2,\dots,\x_N]^T, where \X\in \mathbb{R}^{N\times H}
	\end{equation*}{}
	Then, uni-variate ranking function $f_{\theta}(\x)$ and multivariant scoring function $\mathcal{F}_{\theta}(\X)$ for a ranked list can be defined as:
	\begin{equation*}
		\begin{split}
			f_{\theta}(\X)&=[f_{\theta}(\x_1),f_{\theta}(\x_2),\dots,f_{\theta}(\x_N)] \\
			\mathcal{F}_{\theta}(\X)&=[\mathcal{F}^1_{\theta}(\X),\mathcal{F}^2_{\theta}(\X)^,\dots, \mathcal{F}_{\theta}^N(\X)]
		\end{split}
	\end{equation*}{}
	where $\theta$ is the parameter. The main difference is that multivariant scoring functions take the whole list as input, while uni-variate scoring functions only score one document a time.
	\vspace{-0.3cm}
	\subsection{AutoULTR Framework}
	In this paper, we adopt DLA\cite{ai2018unbiased}, an AutoULTR framework which treats the problem of learning a propensity model from click data  (i.e., the estimation of bias in clicks) as a dual problem of constructing an unbiased learning-to-rank model. 
	Formally, let $o_q$, $r_q$ and $c_q$ be the sets of Bernoulli variables that represent whether a document in $\pi_q$ is observed, perceived as relevant, and clicked by a user, respectively. 
	In DLA, an unbiased ranking system $S$ and a propensity model $E$ can be jointly learned by optimizing the local AttRank losses\cite{ai2018learning} as
	\begin{equation} 
	\begin{split}
	{l}(E,q)&=-\sum_{i=1}^{i=|\pi_q|}o_{q}^i \times \log \mathcal{G}^i_{\phi}(\pi_q) \\
	{l}(S,q)&=-\sum_{i=1}^{i=|\pi_q|}r_{q}^i \times  \log  \mathcal{F}^i_{\theta}(\X_q)
	\end{split}{}
	\label{eq:idea}
	\end{equation}{}
	\vspace{-0.3cm}
	\begin{equation} 
	\begin{split}
	\sum_{i=1}^{i=|\pi_q|}\mathcal{G}^i_{\phi}(\pi_q)=1,\sum_{i=1}^{i=|\pi_q|}\mathcal{F}^i_{\theta}(\X_q)=1
	\end{split}{}
	\label{eq:summation_const}
	\end{equation}{}
	where $\mathcal{G}_{\phi}$ and $\mathcal{F_{\theta}}$, parameterized by $\phi$ and $\theta$, compute the propensity scores and relevance scores of each document in the ranked list with a softmax function constrained by Eq.(\ref{eq:summation_const}). 
	To compute ${l}(S,q)$ and ${l}(E,q)$, we need to know the actual relevance (i.e. $r_q^i$) and observation (i.e., $o_q^i$) information of each document. 
	However, in practice, the only data we can get is click (i.e. $c_q$), and $r_q$ and $o_q$ are unknown latent variables. In order to deal with clicks, a common assumption used by most studies is
	\begin{equation}
	P(c^i_q=1)= P(o^i_q=1)P(r^i_q=1)
	\label{eq:click}
	\end{equation}{}
	which means that users click a search result ($c^i_q=1$) only when it is both
	observed ($o^i_q=1$) and perceived as relevant ($r^i_q=1$), and $o^i_q$ and $r^i_q$ are independent to each other.
	With this assumption, unbiased estimation of $l(S,q)$ and $l(E,q)$ can be achieved through inverse propensity weighting (IPW)~\cite{joachims2017unbiased} and inverse relevance weighting (IRW)~\cite{ai2018unbiased} as
	\begin{equation} 
	\begin{split}
	\widetilde{l}_{IRW}(E,q)&
	=-\sum_{i=1,c_{q}^i=1}^{i=|\pi_q|}\frac{\mathcal{F}^1_{\theta}(\X_q)}{\mathcal{F}^i_{\theta}(\X_q)}\times \log \mathcal{G}^i_{\phi}(\pi_q)
	\\
	\widetilde{l}_{IPW}(S,q)&=-\sum_{i=1,c_{q}^i=1}^{i=|\pi_q|}\frac{\mathcal{G}^1_{\phi}(\pi_q)}{\mathcal{G}^i_{\phi}(\pi_q)}\times \log \mathcal{F}^i_{\theta}(\X_q)
	\end{split}{}
	\label{eq:unbiased}
	\end{equation}{}
	\vspace{-0.3cm}
	\begin{equation*}
		\E_{r_q}[\widetilde{l}_{IRW}(E,q)]\stackrel{\Delta}{=} l(E,q),
		\E_{o_q}[\widetilde{l}_{IPW}(S,q)]\stackrel{\Delta}{=} l(S,q),
	\end{equation*}{}
	where $\stackrel{\Delta}{=}$ means equal or linearly correlated with a positive constant factor. Then the final optimization losses in DLA are
	\begin{equation}
	\widetilde{\mathcal{L}}(S)=\sum_{q\in Q}\widetilde{l}_{IPW}(S,q),~~~~~~ \widetilde{\mathcal{L}}(E)=\sum_{q\in Q}\widetilde{l}_{IRW}(E,q)
	\label{eq:loss}
	\end{equation}{}
	where $Q$ is the set of all possible queries. During training, we update $\theta$ and $\phi$ with the derivatives of $\widetilde{l}_{IPW}(S,q)$ and $\widetilde{l}_{IRW}(E,q)$ respectively and repeat the process until the algorithm converges.
	\vspace{-0.3cm}
	\subsection{Convergence Analysis}
	\vspace{-0.1cm}
	In this section, we analyze the compatibility of multivariate scoring functions and DLA in theory.
	Firstly, we give definition of permutation invariance as:
	\begin{definition}
		Let $S_n$ be the set of all permutations of indices $\{1,\dots,n\}$, A function $f$: $X^n\xrightarrow{} Y^n$ is permutation invariant~\cite{lee2019set} iff for any permutation function $\Pi$, $f(\Pi(X))=\Pi(f(X))$ , i.e.,
		\begin{equation*}
			f(x_{\Pi(1)},\dots,x_{\Pi(n)})=[f^{\Pi(1)}(X),\dots,f^{\Pi(n)}(X)]
			\label{def:permutation_inva}
		\end{equation*}{}
		where $X=[x_1,\dots,x_n]$, $f^{\Pi(i)}(X)$ is the $\Pi(i)\text{-} th$ dimension of $f(X)$.
		\label{def:permu_inva}
		\vspace{-15pt}
	\end{definition}
	
	For simplicity, we consider position bias~\cite{joachims2017accurately} as the only bias in click data.
	Then we have $\mathcal{G}^i_{\phi}=\mathcal{G}^i_{\phi}(\pi_q)$, which means propensity is independent with query.
	Let $\Pi(i)=j$ mean putting $\x_j$ on $i\text{-}th$ position of permutated matrix $\Pi(\X)$.
	In theory, DLA will converge when
	\begin{equation}
	\begin{split}
	\frac{\partial \widetilde{ \mathcal{L}} (E)}{\partial \mathcal{G}^i_{\phi}}=0 \implies    \frac{\mathcal{G}^1_{\phi}}{\mathcal{G}^i_{\phi}}&=\frac{\frac{\E[c^1_{q}]}{\E[c^i_{q}]}}{\E[\frac{\mathcal{F}^1_{\theta}(\X_q)}{\mathcal{F}^i_{\theta}(\X_q)}]}=\frac{\frac{\E[r^1_{q}]}{\E[r^i_{q}]}}{\E[\frac{\mathcal{F}^1_{\theta}(\X_q)}{\mathcal{F}^i_{\theta}(\X_q)}]} \frac{\E[o^1]}{\E[o^i]}
	\end{split}{}
	\label{eq:convergence}
	\end{equation}{}
	Considering any permutation funciton $\Pi$, the original  $i\text{-}{th}$ document in $\pi_q$ is in the  $\Pi^{-1}(i)\text{-} th$ document in the permuted list, where $\Pi^{-1}$ is the inverse function of $\Pi$. Note that in the following analysis, the default ranking of documents is original ranking $\pi_q$ shown to users if not explicitly pointed out. For a permutated ranking, we have
	\begin{equation}
	\begin{split}
	\frac{\mathcal{G}^{\Pi^{-1}(1)}_{\phi}}{\mathcal{G}^{\Pi^{-1}(i)}_{\phi}}&=\frac{\frac{\E[\Pi(r_{q})^{\Pi^{-1}(1)}]}{\E[\Pi(r_{q})^{\Pi^{-1}(i)}]}}{\E[\frac{\mathcal{F}_{\theta}^{\Pi^{-1}(1)}(\Pi(\X_q))}{\mathcal{F}_{\theta}^{\Pi^{-1}(i)}(\Pi(\X_q))}]} \frac{\E[o^{\Pi^{-1}(1)}]}{\E[o^{\Pi^{-1}(i)}]}
	\end{split}{}
	\label{eq:dual_learning}
	\end{equation}{}
	\vspace{-10pt}
	
	\subsubsection{Necessary Condition}
	When DLA converges and propensity is correctly estimated, we have
	\begin{equation}
	\begin{split}
	\forall \Pi,\forall i,~~~~ \frac{\mathcal{G}^{\Pi^{-1}(1)}_{\phi}}{\mathcal{G}^{\Pi^{-1}(i)}_{\phi}}&= \frac{\E[o^{\Pi^{-1}(1)}]}{\E[o^{\Pi^{-1}(i)}]},\frac{\mathcal{G}^1_{\phi}}{\mathcal{G}^i_{\phi}}=\frac{\E[o^1]}{\E[o^i]}
	\end{split}{}
	\label{eq:proensity}
	\end{equation}{}
	Assuming that the relevance of a document $d_i$ would not change after moving to a different position, then we have 
	\begin{equation}
	\Pi(r_{q})^{\Pi^{-1}(i)}=r^i_{q}
	\label{eq:relevance_permu}
	\end{equation}{}
	Considering \Cref{eq:summation_const,eq:relevance_permu,eq:proensity,eq:dual_learning,eq:convergence}, we can get
	\begin{equation}
	\frac{\mathcal{F}^1_{\theta}(\X)}{\mathcal{F}^i_{\theta}(\X)}=\frac{\mathcal{F}_{\theta}^{\Pi^{-1}(1)}(\Pi(\X))}{\mathcal{F}_{\theta}^{\Pi^{-1}(i)}(\Pi(\X))}\iff \mathcal{F}^i_{\theta}(\X)=\mathcal{F}_{\theta}^{\Pi^{-1}(i)}(\Pi(\X))
	\label{eq:mathcal_F}
	\end{equation}{}
	Then, we insert $i=\Pi(j)$ in Eq.\ref{eq:mathcal_F}, and we have
	\begin{equation}
	\forall \Pi,\forall j, ~~~~\mathcal{F}_{\theta}^{j}(\Pi(\X))=  \mathcal{F}_{\theta}^{\Pi(j)}(\X)
	\label{eq:permu_capitalF}
	\end{equation}{}
	which means that $\mathcal{F}_{\theta}$ is permutation invariant according to Definition \ref{def:permu_inva}.
	This indicates that permutation invariance is a necessary condition for the convergence of DLA.
	
	\vspace{-0.2cm}
	\subsubsection{Sufficient Condition}
	
	Suppose that $\mathcal{F}_{\theta}$ is permutation invariant, then the estimated relevance score for a document from $\mathcal{F}_{\theta}$ is independent of its position. 
	Because $\mathcal{G}_{\phi}$ only takes the positions as input, $\mathcal{F}_{\theta}$ and $\mathcal{G}_{\phi}$ are independent to each other and can separately estimate the relevance and propensity during training.
	As proven by Ai et al.~\cite{ai2018unbiased}, DLA is guaranteed to converge in this case, which means that permutation invariance can be a sufficient condition for the convergence of DLA.

	
	
	
	
	\section{Experiment}
	So far, we have proven that permutation invariance is the sufficient and necessary condition for learning-to-rank models to converge in DLA in theory.
	In this section, we describe our experiments on two large-scale benchmarks for further demonstrations.
	\vspace{-0.3cm}
	\subsection{Simulation Experiment Settings}
	\vspace{-0.1cm}
	\subsubsection{Datasets and Click Simulation}
	To fully explore the performance of multivariate scoring functions in AutoULTR, we conducted experiments on Yahoo! LETOR set 1\footnote{\url{https://webscope.sandbox.yahoo.com}},and Istella-S LETOR \footnote{\url{http://blog.istella.it/istella-learning-to-rank-dataset/}} with derived click data.
	Similar to previous studies~\cite{joachims2017unbiased}, we trained a SVM$^{rank}$ model\footnote{\url{http://www.cs.cornell.edu/people/tj/svm_light/svm_rank.html}} (which we refer to as \textit{Prod.})
	using 1\% of the training data with real relevance judgements to generate the original ranked list $\pi_q$. We then sampled clicks ($c_q^i$) on documents according to Eq.~(\ref{eq:click}) with $o_q^i$, and $r_q^i$ as,
	\begin{equation*}
		\begin{split}
			P(o_q^i=1|\pi_q)&=P(o_i=1)=\rho_i\\
			P(r_q^i=1|\pi_q)&=\epsilon+(1-\epsilon)\frac{2^y-1}{2^{y_{max}}-1}
		\end{split}{}
	\end{equation*}{}
	where $\rho$ is acquired through eye-traching experiments~\cite{joachims2017accurately}, $y\in [0,4]$ is the 5-level relevance label in both datasets where $y_{max}=4$, and $\epsilon$ is used to model click noise so that irrelevant document($y=0$) can also be clicked. For simplicity, we fixed the value of $\epsilon$ as 0.1.
	\vspace{-0.2cm}
	\subsubsection{Models and Evaluation Measures}
	In this paper, we focus on two state-of-art multivariate scoring functions.
	The first one is DLCM\cite{ai2018learning}, a RNN model with gated recurrent unit(GRU) that treats the final network state as context to score each document.
	It is permutation variant by nature. 
	The second one is SetRank\cite{pang2020setrank}, constructed with multi-head self-attention networks, which is permutation invariant. 
	For DLCM, we adopt three kinds of input orders, namely the original ranking of documents (i.e., DLCM$^{init}$) created by Prod.; the reverse of the original ranking (i.e., DLCM$^{rever}$), and a random permutation of the original ranking (i.e., DLCM$^{rand}$).
	For comparison, we include a uni-variate scoring function based on deep neural networks (DNN) as our baseline.
	We also include a DNN model that directly use clicks as relevance labels, which is referred to as DNN$^{navie}$.
	The source code can be found here\footnote{\url{https://github.com/Taosheng-ty/CIKM_2020_Multivariate_AutoULTR.git}}.
	
	
	
	All models were tuned and selected based on their performances
	on the validation set according to $nDCG@10$. Each model was trained for five times and reported the mean performance
	The batch size was set to be 64. 
	Learning rate was tuned between 0.1 and 0.01, and we stoped training after 60k steps. 
	During training, we set the size of ranked list as 10, while during validating and testing, we tested our ranking model on all documents to each query. 
	We reported both ERR and nDCG metrics at ranks of 3 and 10 to show ranking performance. 
	Besides, in order to show performance of propensity estimation,
	we computed the mean square error (MSE) between the true inverse propensity weights $(\rho_1/\rho_i )$ and the estimated inverse propensity
	weights $(\mathcal{G}^1/\mathcal{G}^i)$ as 
	\begin{equation*}
		MSE_{propen}=\frac{1}{|\pi_q|}\sum_{i=1}^{|\pi_q|}(\frac{\mathcal{G}^1}{\mathcal{G}^i}-\frac{\rho_1}{\rho_i})^2
		\vspace{-0.4cm}
	\end{equation*}{}
	
	\begin{table}
		\centering
		\caption{Performance comparison of different scoring functions in AutoULTR. Significant improvements or degradations with respect to DNN are indicated with $\p/\m$ ($p<0.05$).}
		\vspace{-0.4cm}
		\scalebox{0.75}{
			\begin{tabular}{c|c|c|c|c|c} \hline
				Scoring Model&ERR @3 & nDCG @3 & ERR @10&nDCG @10& $MSE_{propen}$ \\\hline
				\multicolumn{6}{c}{(a) Yahoo!} \\\hline
				DLCM$^{rever}$ & 0.414$^\m$ &0.665$^\m$ &0.452$^\m$ &0.739$^\m$ &7.40$^\m$ \\\hline
				DLCM$^{init}$    & 0.427 &0.686$^\m$ &0.464 &0.756$^\m$ &9.31$^\m$\\\hline
				DLCM$^{rand} $ &0.425$^\m$& 0.680$^\m$&0.462$^\m$&0.752$^\m$&0.015\\\hline
				SetRank  & 0.428& 0.694$^\p$& 0.464 &0.762$^\p$&0.097\\\hline \hline
				DNN   &0.427&0.692  &0.464&0.760&0.048\\\hline
				
				DNN$^{naive}$ & 0.411$^\m$&0.664$^\m$& 0.449 $^\m$& 0.740 $^\m$&-\\\hline \hline
				Prod.  & 0.374$^\m$ & 0.611$^\m$ & 0.416$^\m$ & 0.705$^\m$&-\\\hline

				\multicolumn{6}{c}{(b) Istella-s} \\\hline
				
				DLCM$^{rever}$ & 0.676$^\m$ &0.601$^\m$ &0.703$^\m$&0.695$^\m$ &17.5$^\m$\\\hline
				DLCM$^{init}$ & 0.700 &0.629&0.724&0.707$^\m$ &16.1$^\m$\\\hline
				DLCM$^{rand} $ &0.690$^\m$ &0.620$^\m$&0.714$^\m$&0.710$^\m$&0.023\\\hline
				SetRank  & 0.706& 0.636& 0.730 &0.721$^\p$&0.135$^\m$\\\hline \hline
				DNN   &0.704 & 0.633& 0.727 &0.716&0.033\\\hline      
				DNN$^{naive}$ & 0.683$^\m$ &0.610$^\m$ & 0.709$^\m$ & 0.700$^\m$&-\\\hline \hline
				Prod.  &0.640$^\m$ & 0.562$^\m$ & 0.669$^\m$ & 0.663$^\m$&-\\\hline
		\end{tabular} }
		\label{tab:results}
	\end{table}{}
	\begin{figure}
		\vspace{-0.42cm}
		\captionsetup[table]{font=small,skip=0pt}
		\centering
		\includegraphics[scale=0.46]{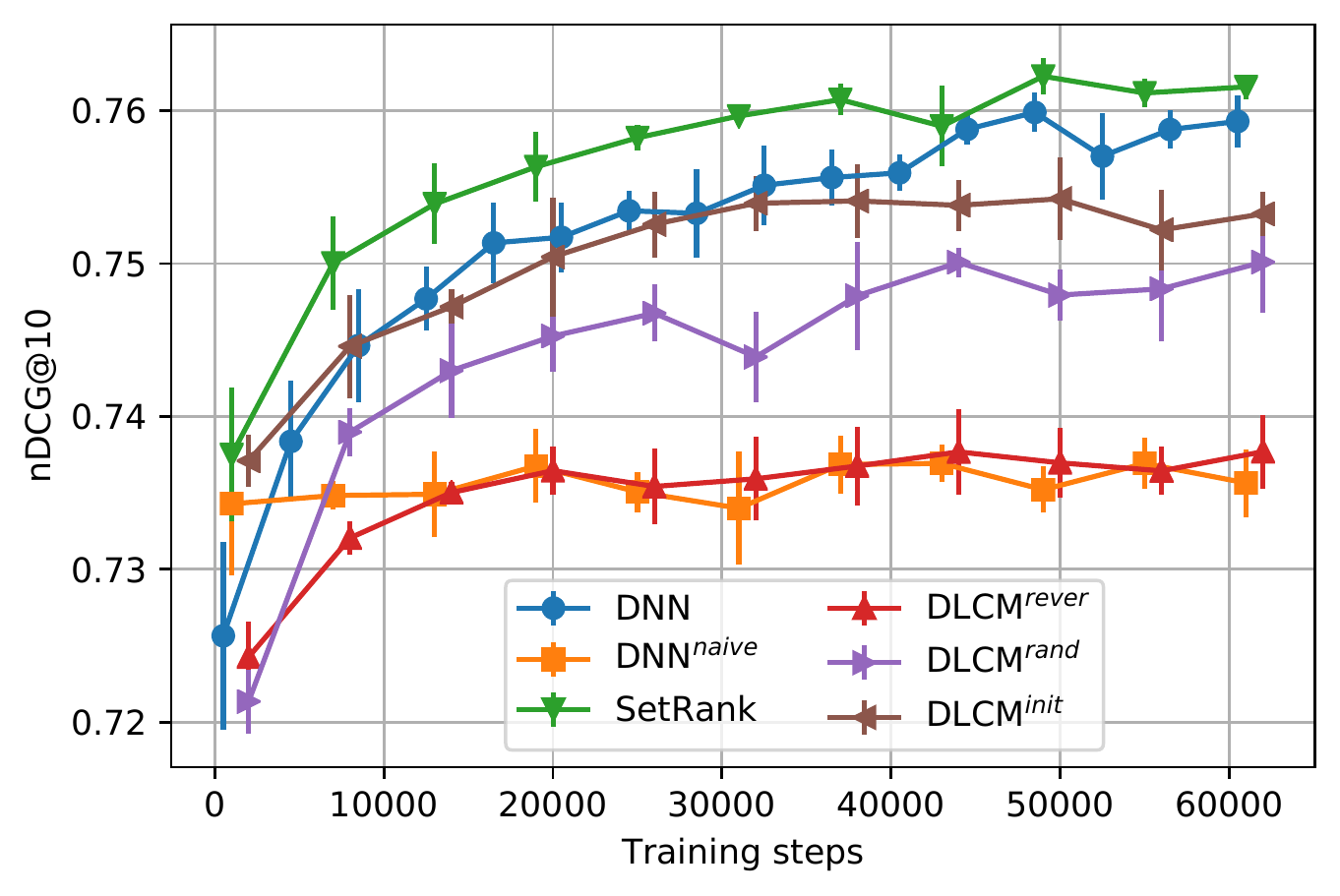}
		\vspace{-0.55cm}
		\caption{Test performance on Yahoo! LETOR set 1.}
		\label{fig:yahoo}
	\end{figure}{}
	\begin{figure}
		\vspace{-0.53cm}
		\centering
		\includegraphics[scale=0.46]{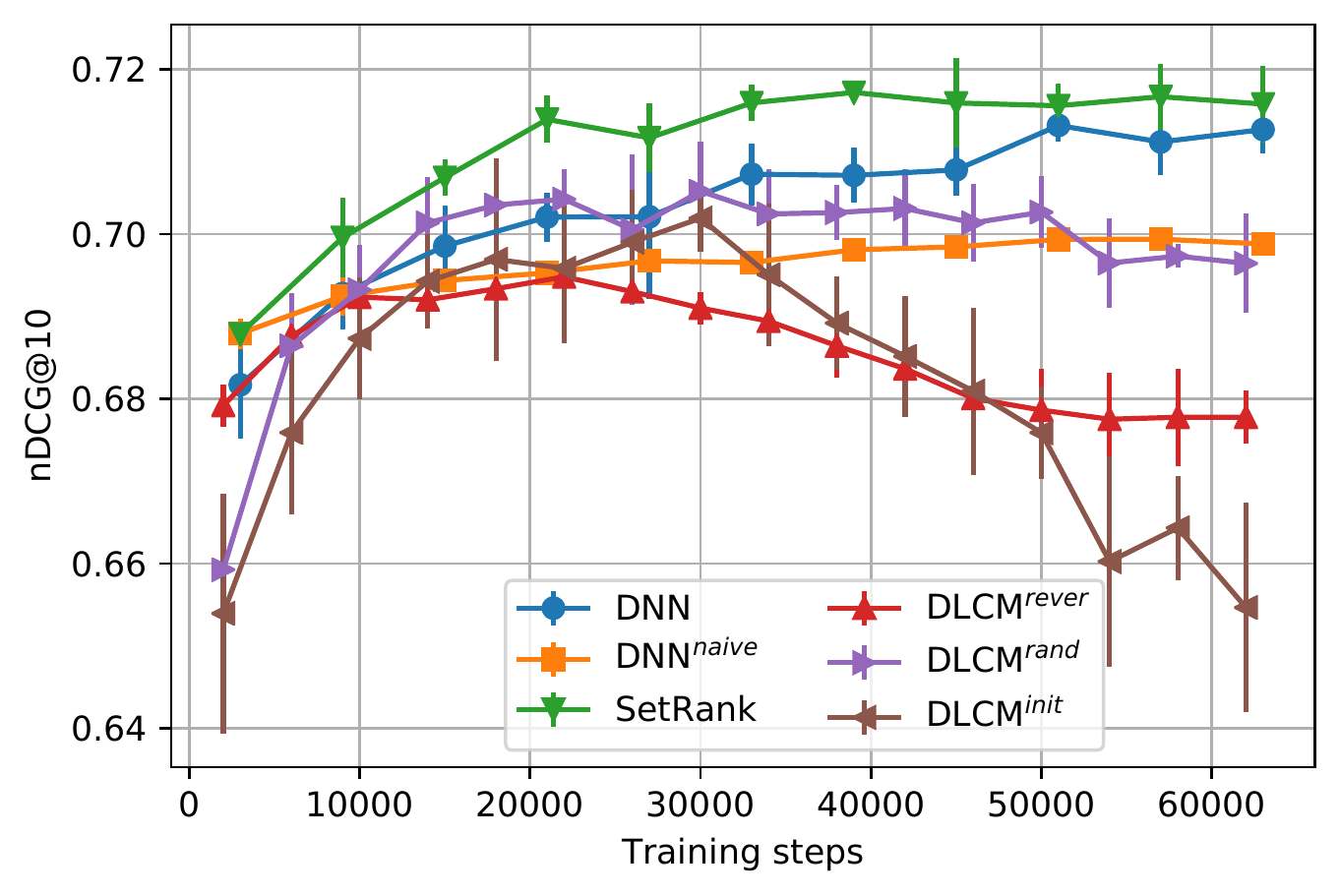}
		\vspace{-0.55cm}
		\caption{Test performance on Istella-s.}
		\label{fig:istella-s}
		\vspace{-0.6cm}
	\end{figure}{}
	
	
	
	
	\subsection{Experimental Results and Analysis}
	\vspace{-0.05cm}
	A summary of the results are shown in Table \ref{tab:results}, Fig.\ref{fig:yahoo} and Fig.\ref{fig:istella-s}. 
	\vspace{-0.2cm}
	\subsubsection{Permutation variant and invariant ranking model.} As we can see from Table \ref{tab:results}, DNN and SetRank, two permutation invariant ranking models, work well with DLA and outperform DNN$^{naive}$, which directly trains DNN with click data. 
	However,  DLCM$^{init}$ and DLCM$^{rever}$ show terrible performance on estimating propensity (i.e., high MSE) and poor ranking performance when compared to DNN and SetRank. 
	The results indicate that AutoULTR with permutation variant functions is not guaranteed to converge and get an unbiased ranker. 
	\vspace{-0.2cm}
	\subsubsection{Comparison between uni-variant and multivariant ranking model.}
	Here we only consider SetRank and DNN, which are multivariant and uni-variant respectively.
	Both SetRank and DNN are permutation invariant and theoretically principled to converge with AutoULTR. 
	From Fig.~\ref{fig:yahoo} and Fig.~\ref{fig:istella-s}, we can see that the ranking performance of SetRank significantly outperforms DNN, especially on Istella-S LETOR. 
	The results confirm the arguments from previous studies that multivariate scoring functions are superior to uni-variate ones because the former can capture the contextual information and cross-document interactions.
	\vspace{-0.2cm}
	\subsubsection{Comparison of DLCM with different input order.}\label{sec:dlcm} 
	We noticed that DLCM$^{rand}$ could get a perfect estimation of the propensity but failed to get ranking performance as good as SetRank and DNN. We think the reason might be that random input order makes permutation variant models hard to remember any pattern in order, thus empirically achieve permutation invariance. As for ranking performance, we can interpret it from the RNN structure. In DLCM, the final input has the most impact on the final network state, which is viewed as context to help score each document. Random input sequence results in a random context which would make DLCM$^{rand}$ hard to score documents.
	This could also explain the bad performance of DLCM$^{rever}$ as it always takes documents in the reverse order of the original ranking produced by Prod.
	\vspace{-0.1cm}
	\section{CONCLUSION AND FUTURE WORK}
	\vspace{-0.1cm}
	In this work, we explore the potential of multivariate scoring functions in AutoULTR. Based on existing AutoULTR algorithms, We prove that permutation invariance is a crucial factor for a multivariate scoring function to be included in AutoULTR. With two existing multivariate functions and one AutoULTR algorithm, we conduct experiments based on two benchmark datasets, the results of which align with our theoretical analysis. AutoULTR models with permutation-invariant multivariate scoring functions significantly outperform those with uni-variate scoring functions and permutation-variant multivariate scoring functions.
	Our work represents an initial attempt to include multivariate scoring in AutoULTR. In the future, we may base on our analysis to propose novel multivariate scoring functions for AutoULTR.

	\vspace{-0.2cm}
	\begin{acks}
		\vspace{-0.1cm}
		This work was supported in part by the School of Computing,
		University of Utah. Any opinions, findings and conclusions or
		recommendations expressed in this material are those of the authors
		and do not necessarily reflect those of the sponsor.
	\end{acks}
	\vspace{-0.2cm}
	\bibliographystyle{ACM-Reference-Format}
	\bibliography{sample-sigconf}
\end{document}